\documentclass[final,3p,times,twocolumn]{elsarticle}
 \biboptions{comma,sort&compress}
\usepackage{here}
 \usepackage{graphicx}
  \usepackage{epsfig}

\newcommand{\rd}{{\rm d}}

\def\beq {\begin{equation}}
\def\eeq {\end{equation}}
\def\bea {\begin{eqnarray}}
\def\eea {\end{eqnarray}}

\def\nn {\nonumber}

\def\bfl {\mbox{\boldmath $\lambda$}}

\def\tauKpi{\tau\to K \pi\nu_\tau}

\journal{Nuc. Phys. (Proc. Suppl.)}

\begin{document}

\begin{frontmatter}



\title{Constraining the $K\pi$ vector form factor by $\tau\to K\pi \nu_\tau$ and $K_{l3}$ decay data} 

 \author[label1]{Diogo R. Boito\corref{cor1}}
  \address[label1]{Grup de F\'{\i}sica Te\`orica and IFAE, Universitat Aut\`onoma de Barcelona,
E-08193 Bellaterra (Barcelona), Spain\\}
\cortext[cor1]{Speaker}
\ead{boito@ifae.es}

 \author[label1]{Rafel Escribano}
\ead{rescriba@ifae.es}
\author[label2]{Matthias Jamin}
  \address[label2]{Instituci\'o Catalana de Recerca i Estudis Avan\c cats (ICREA),
IFAE and Grup de F\'{\i}sica Te\`orica, Universitat Aut\`onoma de
Barcelona,\\
E-08193 Bellaterra (Barcelona), Spain}
\ead{jamin@ifae.es}


\begin{abstract}
\noindent
 A subtracted dispersive representation of the $K\pi$ vector form
 factor, $F_+^{K\pi}$, is used to fit the Belle spectrum of $\tauKpi$ decays
 incorporating constraints from  results on $K_{l3}$
 decays. Through the use of three subtractions, the slope and
 curvature of  $F_+^{K\pi}$ are obtained directly from the
 data yielding $\lambda_+'=(25.49 \pm 0.31) \times 10^{-3}$ and
 $\lambda_+''= (12.22 \pm 0.14) \times 10^{-4}$. The phase-space
 integrals relevant for $K_{l3}$  analyses are
 calculated. Additionally, from  the pole position on the second Riemann sheet 
 the mass and width of the $K^*(892)^\pm$ are found to be $m_{K^*(892)^\pm}= 892.0\pm 0.5\, \,\mbox{MeV}$ and $\Gamma_{K^*(892)^\pm}= 46.5 \pm1.1 \,\, \mbox{MeV}$. Finally, we  study the $P$-wave isospin-1/2 $K\pi$   phase-shift and its threshold
 parameters.
\end{abstract}


\end{frontmatter}


\section{Introduction}

The non-perturbative physics of $K\to \pi\, l\, \nu_l$ ($K_{l3}$) and
$\tauKpi$ decays is governed by two Lorentz-invariant $K\pi$ form
factors, namely the vector, denoted $F^{K\pi}_+(q^2)$, and the scalar,
$F^{K\pi}_0(q^2)$.  A good knowledge of these form factors paves the
way for the determination of many parameters of the Standard Model,
such as the quark-mixing matrix element $|V_{us}|$ obtained from
$K_{l3}$ decays~\cite{LR84}, or the strange-quark mass $m_s$
determined from the scalar QCD strange spectral function~\cite{JOP4}.

Until recently, the main source of experimental information on $K\pi$
form factors have been $K_{l3}$ decays. Latterly, five experiments
have collected data on semileptonic and leptonic $K$ decays: BNL-E865,
KLOE, KTeV, ISTRA+, and NA48. The results from these analyses yielded
an important amount of information on the form factors as well as
stringent tests of QCD at low-energies and of the Standard Model
itself (for a recent review on theoretical and experimental aspects of
kaon physics we refer to Ref.~\cite{Antonelli10}).  Additional
knowledge on the $K\pi$ form factors can be gained from the dominant
Cabibbo-suppressed $\tau$ decay: the channel $\tauKpi$.  Presently, the
$B$ factories have become a superior source of high-statistics data
for this reaction by virtue of the important cross-section for
$e^+e^-\to \tau^+\tau^-$ around the $\Upsilon(4S)$ peak.  A detailed
spectrum for $\tau \to K_S\,\pi^-\nu_\tau$ produced and analysed by
Belle was published in 2007~\cite{Belle}. Also, preliminary BaBar
spectra with similar statistics have appeared recently in conference
proceedings~\cite{Babar} and, finally, BESIII should produce results
for this decay in the future~\cite{BESIII}.  The new data sets provide
the substrate for up-to-date theoretical analyses of the $K\pi$ form
factors. In Ref.~\cite{BEJ} we have performed a reanalysis of the
$\tauKpi$ spectrum of~\cite{Belle}. More recently, we carried out an
analysis with restrictions from $K_{l3}$ experiments~\cite{BEJ2010}.

On the theory side, the knowledge of these form factors consists of
two tasks. The first of them is to determine their value at the
origin, $F_{+,0}(0)$,  crucial in order to disentangle
the product $|V_{us}|F_{+,0}(0)$. Historically, chiral perturbation
theory has been the main tool to study $F_{+,0}(0)$, but recently
lattice QCD collaborations have produced more accurate results for
this quantity~\cite{Lattice}. Second, one must know the energy
dependence of the form factors, which is required when calculating
phase-space integrals for $K_{l3}$ decays or when analysing the
detailed shape of the $\tauKpi$ spectrum. In our work we concentrate
on the latter aspect of the problem and therefore it is convenient to
introduce form factors normalised to one at the origin
\beq
\tilde F_{+,0}(q^2)= F_{+,0}(q^2)/F_{+,0}(0).
\eeq

A salient feature of the form factors in
 the kinematical region relevant for $K_{l3}$ decays, {\emph i.e.}
 $m_l^2<q^2<(m_K-m_\pi)^2$, is that they are real.  Within the allowed
 phase-space they admit a Taylor expansion and the energy
 dependence is customarily translated into constants
 $\lambda_{+,0}^{(n)}$ defined as\footnote{From now on we refrain from writing the superscript
$K\pi$ on the form factors.}
\beq
\tilde F_{+,0}(q^2)=1 + \lambda_{+,0}' \frac{q^2}{m_{\pi^-}^2} + \frac{1}{2}\lambda_{+,0}'' \left( \frac{q^2}{m_{\pi^-}^2}\right)^2 + \cdots \,.
\label{FFTaylor}
\eeq
In $\tauKpi$ decays, however, since $(m_K+m_\pi)^2<q^2<m_\tau^2$, one
deals with a different kinematical regime in which the form factors
develop imaginary parts, rendering the expansion of
Eq.~(\ref{FFTaylor}) inadmissible.  One must then resort to more
sophisticated treatments. Moreover, in order to fully benefit from the available
experimental data, it is desirable to employ representations of the form factors
that are valid for both $K_{l3}$ and $\tauKpi$ decays.  Dispersive representations
of the form factors provide a powerful tool to achieve this goal.

From general principles, the form factors must satisfy a dispersion
relation. Supplementing this constraint with unitarity, the dispersion
relation has a well-known closed-form solution within the elastic
approximation referred to as the Omn\`es
representation~\cite{Omnes}. Although simple, this solution requires
the detailed knowledge of the phase of $F_+(s)$ up to infinity, which
 is unrealistic. An advantageous strategy to circumvent this
problem is the use of additional  subtractions, as done, for instance, for the pion form factor in Ref.~\cite{PP2001}. Subtractions in the dispersion relation entail a
suppression of the integrand in the dispersion integral for higher
energies. An $n$-times-subtracted form factor exhibits a suppression
of $s^{-(n+1)}$ in the integrand. Thereby, the information that was
previously contained in the high-energy part of the integral is
translated into $n-1$ subtraction constants. In Ref.~\cite{BEJ} we
performed fits to the Belle spectrum of $\tauKpi$ varying the number
of subtractions and testing  descriptions with one and two vector
resonances. The outcome of these tests, described in detail in
Ref.~\cite{BEJ}, is that for our purposes an optimal description of
$F_+(s)$ is reached with three subtractions and two resonances. Here
we quote the resulting  expression
\bea
\tilde F_+(s) \,=\, \exp\Bigg[ \alpha_1\, \frac{s}{m_{\pi^-}^2} +
\frac{1}{2}\alpha_2\frac{s^2}{m_{\pi^-}^4}  \nn \\ + \frac{s^3}{\!\pi}
\int\limits^{s_{\rm cut}}_{s_{K\pi}} \!\!ds'\, \frac{\delta(s')}
{(s')^3(s'-s-i0) }\Bigg] \,.\label{dispFF}
\eea
In the last equation, $s_{K\pi}=(m_{K^0}+m_{\pi^-})^2$ and the two
subtraction constants $\alpha_1$ and $\alpha_2$ are related to the
Taylor expansion of Eq.~(\ref{FFTaylor}) as $\lambda_+'=\alpha_1$ and
$\lambda_+''=\alpha_2+\alpha_1^2$.  It is opportune to treat them as
free parameters that capture our ignorance of the higher energy part
of the integral.  The constants $\lambda_+'$ and $\lambda_+''$ can
then be determined through the fit. The main advantage of this
procedure, advocated for example in
Refs.~\cite{PP2001,Bernardetal,BEJ}, is that the
subtraction constants turn out to be less model dependent as they are
determined by the best fit to the data. It is important to stress that
Eq.~(\ref{dispFF}) remains valid beyond the elastic approximation
provided $\delta(s)$ is the phase of the form factor, instead of the
corresponding scattering phase. But, of course, in order to employ it
in practice we must have a model for the phase. As described in detail
in Ref.~\cite{BEJ}, we take a form inspired by the RChT treatment
of~Refs.~\cite{JPP20062008} with two vector resonances.  For the
detailed expressions we refer to the original works. With
Eq.~(\ref{dispFF}), the transition from the kinematical region of
$\tauKpi$ to that of $K_{l3}$ decays is straightforward and the
dominant low-energy behaviour of $F_+(s)$ is encoded in $\lambda_+'$
and $\lambda_+''$. The cut-off $s_{\rm cut}$ in the dispersion
integral is introduced to quantify the suppression of the higher
energy part of the integrand. The stability of the results is checked
varying this cut-off in a wide range from $1.8\, \mbox{GeV} < s_{\rm
  cut} < \infty$. As a final comment, since $\alpha_{1,2}$ are
determined by the data, in the limit $s\to \infty$ the asymptotic
behaviour of $F_+(s)$ cannot be satisfied. This is so because a
perfect cancellation between terms containing $\alpha_{1}$ and
$\alpha_{2}$ with polynomial terms coming from the dispersion integral
must occur in order to guarantee that $F_+(s)$ vanishes as $1/s$. We
have checked that our form factor, within the entire range where we
apply it (and beyond), is indeed a decreasing function of $s$ which
renders our approach credible.

In $\tauKpi$ decays, the scalar form factor is suppressed
kinematically. Albeit marginal, the contribution from $F_0$ cannot be
neglected in the lower energy part of the spectrum. Here, we keep this
contribution fixed using the results for $F_0$ from the coupled-channel
dispersive analysis of Refs.~\cite{JOP1,JOP4}.

\section{Fits to $\mathbf{\tauKpi}$  with constraints from $\mathbf{K_{l3}}$  }

The analysis of the spectrum for $\tauKpi$ produces a wealth of
physical results, many of them with great accuracy, e.g., the mass
and width of the $K^*(892)$. We have advocated by means of Monte Carlo
simulations that a joined analysis of $\tauKpi$ and $K_{l3}$ spectra
further constrains the low-energy part of the vector form-factor
yielding results with a better precision~\cite{BEJ}. This idea was pursued in our
recent work~\cite{BEJ2010}.

In order to include the experimental information available from
$K_{l3}$ decays---and for the want of true unfolded data sets from
these experiments---we adopt the following strategy. In our fits,
the $\chi^2$ that is to be minimised contains a standard part from the
$\tauKpi$ spectrum and a piece which constrains the parameters
$\lambda_+^{(' ,'')}$ using information from $K_{l3}$
experiments. This is realized in practice as
\bea
\label{chi2comb}
 \chi^2 = \sum_{i=1}^{90}{}^{\prime} \left(\frac{N_i^{\rm th} - N^{\rm exp}_i}{\sigma_{N^{\rm exp}_i}} \right )^2  + \left(\frac {\bar B_{K\pi} - B_{K\pi}^{\rm exp}}{\sigma_{B^{\rm exp}_{K\pi}}}\right)^2\nn \\
+  (\bfl_+^{\rm th} -\bfl_+^{\rm exp})^{\rm T} V^{-1}  (\bfl_+^{\rm th} -\bfl_+^{\rm exp}) ,
\eea
where the first two terms on the r.h.s. are those of a fit to the
spectrum of $\tauKpi$ and the third one encodes the information from
$K_{l3}$ analyses and acts as a sort of prior (in the Bayesian
sense) for the parameters $\lambda_+'$ and $\lambda_+''$.  In the last
equation the theoretical
number of events $N_i^{\rm th}$ in the $i$-th bin is taken to be (as explained in Ref.~\cite{JPP20062008})
\beq
\label{Nth}
N_i^{\rm th} =\mathcal{N}_T\,  \frac{1}{2}\,\frac{2}{3}  \,\Delta^{i}_{\rm b}\,  \frac{1}{\Gamma_\tau \, \bar B_{K\pi}}  \frac{\rd\Gamma_{K\pi}}{\rd\sqrt{s}} (s_{\rm b}^ i)\,,
\eeq
where $\mathcal{N}_T$ is the  total number of events, the factor $\frac{1}{2}$ and $\frac{2}{3}$ account
for the fact that the $K_S\pi^-$ channel was analysed, $\Delta^{i}_{\rm b}$ is the  width of the $i$-th bin, $\Gamma_\tau$  is the  total $\tau$ decay  width, $\bar
B_{K\pi}$ is a normalisation  constant that, for a perfect description
of the spectrum, should be the $\tauKpi$ branching ratio, and, finally,
$s_{\rm b}^ i $ is the centre of the $i$-th bin.
 Furthermore, $N^{\rm exp}_i$ and $\sigma_{N^{\rm exp}_i}$ are,
respectively, the experimental number of events in the Belle spectrum~\cite{Belle} and the corresponding
uncertainty in the $i$-th bin. The prime in the symbol of sum
indicates that bins 5, 6, and 7 are excluded from the
minimisation\footnote{For a detailed discussion of the fit procedure
  we refer to~\cite{BEJ2010}}.
The second term on the right-hand side of Eq.~(\ref{chi2comb}) introduces
an additional restriction that allows us to treat the normalisation
$\bar B_{K\pi}$ of Eq.~(\ref{Nth}) as a free parameter.

In the last term of  Eq.~(\ref{chi2comb}), the vectors $\bfl_ +^{\rm th, exp}$ are given by
\beq
\bfl_ +^{\rm th, exp}=
\left(\begin{array}{c}
\lambda_+^{\prime\, \rm th, exp} \\
\lambda_+^{\prime\prime\, \rm th, exp}
\end{array}\right)\, ,
\eeq
and the $2\times 2$  matrix $V$ is the experimental covariance for $\bfl_+^{\rm exp}$  such that
\beq
V_{ij}= \rho_{ij}\, \sigma_i\, \sigma_ j\, ,
\eeq
where the indices  refer to $\lambda_+'$ and $\lambda_+''$,
$\rho_{ij}$ is the correlation coefficient ($\rho_{ij}=1$ if $i=j$),
and $\sigma_{i}$ the experimental errors on $\lambda_+'$ and
$\lambda_+''$.  For the experimental values we employ the results of
the compilation of $K_L$ analyses performed by Antonelli {\it et al.} for the FlaviaNet Working Group on Kaon Decays in
Ref.~\cite{Antonelli10}: $\lambda_+^{\prime\, \rm exp}=(24.9\pm1.1)\times 10^{-3}$, 
$\lambda_+^{\prime\prime\, \rm exp}=(16\pm 5)\times 10^{-4}$ and
$\rho_{\lambda_+^{\prime},\lambda_+^{\prime \prime}}=-0.95$.

\subsection{Results}

From the minimisation of the $\chi^2$ of Eq.~(\ref{chi2comb}) a
collection of physical results can be derived. Some of them are
obtained directly from the fit, such as $\lambda_+'$ and $\lambda_+''$
and the mass and width of the $K^*(892)$. With the form factor under
control, one can then obtain other results such as the phase-space
integrals for $K_{l3}$ decays. In order to control the uncertainties
and the consistency of the results one must check the stability of the
fit with respect to the cut-off $s_{\rm cut}$ of
Eq.~(\ref{dispFF}). The detailed tables of Ref.~\cite{BEJ2010} attest
that the results are indeed rather stable, but in some cases a
residual $s_{\rm cut}$ dependence contributes to the final uncertainty
we quote. Here, we present the main results of Ref.~\cite{BEJ2010}. A
careful comparison with other results found in the literature can be
found in that reference.

We start by quoting our final results for the mass and the width of
the $K^*(892)^{\pm}$
\bea\label{massandwidth}
m_{K^*(892)^\pm}= 892.03\pm (0.19)_{\rm stat}\pm (0.44)_{\rm sys}\, \,\mbox{MeV}, \nn\\
\Gamma_{K^*(892)^\pm}= 46.53 \pm (0.38)_{\rm stat}\pm (1.0)_{\rm sys}\,\, \mbox{MeV}\, .
\eea
These results are obtained from the complex pole position on the
second Riemann sheet,  $s_{K^*}$, following the definition
$\sqrt{s_{K^*}}= m_{K^*} -(i/2)\Gamma_{K^*}$.
It is important to stress that the mass and width thus obtained are rather
different from the parameters that enter our description
of the phase of $F_+(s)$. When comparing  results from different works
one must always be sure that the same definition is used in all cases.
In Ref.~\cite{BEJ2010}, we showed that our results are compatible with others {\it provided} the pole position prescription is employed for all the analyses.

The final results for the parameters $\lambda_+'$ and $\lambda_+''$ read
\bea\label{lambdas}
\lambda_+'\times 10^{3}  &=&  25.49\pm (0.30)_{\rm stat} \pm (0.06)_{s_{\rm cut}}\,, \nn \\
\lambda_+''\times 10^{4} &=&  12.22\pm (0.10)_{\rm stat} \pm (0.10)_{s_{\rm cut}}\,.
\eea
In this case, the uncertainty from the variation of $s_{\rm cut}$ contributes as indicated. From the expansion of Eq~(\ref{dispFF}) we can calculate the  third coefficient of a Taylor series of the type of Eq.~(\ref{FFTaylor}). We find 
\beq
\lambda_+'''\times 10^5=8.87\pm (0.08)_{\rm stat} \pm (0.05)_{s_{\rm cut}}\,.
\eeq
These results are in good agreement with other analyses but have
smaller uncertainties since our fits are constrained by $\tauKpi$ and
$K_{l3}$ experiments.

Once the low-energy behaviour of the vector form-factor is obtained
from the fit, we can compare it to the equivalent chiral expansion in
order to determine the low-energy constant $L_9^r$. Using the
$\mathcal{O}(p^4)$ expressions of Ref.~\cite{GL} with $F_0^2=F_\pi^2$
we obtain
\beq
L_9^r(m_{K^*})\big|_{F_0^2=F_\pi^2}\times 10^{3} = 5.19\pm (0.07)_{\rm stat}\, .
\eeq
It is well known that the dominant uncertainty is given by the
truncation of the series at $\mathcal{O}(p^4)$. As an estimate of
$\mathcal{O}(p^6)$ effects we can employ $F_0^2=F_\pi F_K$ which gives
\beq
L_9^r(m_{K^*})\big|_{F_0^2=F_\pi F_K}\times 10^{3} = 6.29\pm (0.08)_{\rm stat}\, .
\eeq

In the extraction of $|V_{us}|$ from the $K_{l3}$ decay widths, one
must perform phase-space integrals where the form-factors play the
central role. The integrals are defined in
Ref.~\cite{LR84,Antonelli10}. From our form-factors we obtain the
following results
\bea
I_{K^0_{e 3}}    =  0.15466(17),  \,\,     I_{K^0_{\mu 3}}  =   0.10276(10),  \nn        \\
I_{K^+_{e 3}}    =  0.15903(17), \,\, I_{K^+_{\mu 3}}  =   0.10575(11). 
\eea
The uncertainties were calculated with a MC sample of parameters
obeying the results of our fits with the correlations properly
included. The final uncertainties are competitive if compared with
the averages of~\cite{Antonelli10} and the central values agree.

Another interesting result that can be extracted from the $\tauKpi$
spectrum is the $K\pi$ isospin-1/2 $P$-wave scattering phase. The decay in
question is indeed a very clean source of information about $K\pi$
interactions, since the hadrons are isolated in the final state. Below inelastic thresholds, the phase of the form-factor is 
the scattering phase, as dictated by Watson's theorem.
The
result of our $P$-wave phase is shown in Fig. 1 where we compare it
with the results of two hadronic experiments~\cite{LASS, Estabrooks}.
\begin{figure}[!ht] 
\includegraphics[width=1\columnwidth,angle=0]{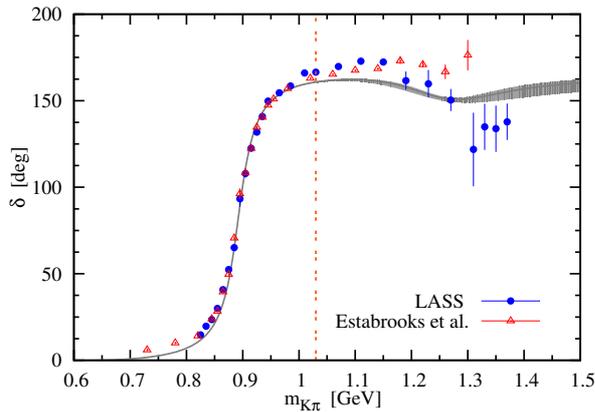}
\caption{{Phase of the form factor $F_+(s)$ and results  from   LASS~\cite{LASS} and  Estabrooks {\it et al.}~\cite{Estabrooks}. The opening of the first inelastic channel, $K^*\pi$, is indicated by the dashed vertical line. The gray band represents the uncertainty due to $s_{\rm cut}$.} } 
\label{Phase}
\end{figure}

From the expansion of the corresponding partial-wave $T$-matrix
element in the vicinity of the $K\pi$ threshold one can determine the
$K\pi$ $P$-wave threshold parameters. With our results, the first three
read
\bea
\label{ScattLengRes}
m_{\pi^-}^3 \, a_1^{1/2} \times 10    &=& 0.166(4), \nn  \\
m_{\pi^-}^5 \, b_1^{1/2} \times 10^{2}&=&   0.258(9), \nn\\
m_{\pi^-}^7\,  c_1^{1/2}\times 10^{3} &=&   0.90(3).   
\eea

\section*{Acknowledgements}

DRB would like to thank the organisers of the conference. We are
grateful to the Belle collaboration in particular to S.~Eidelman,
D.~Epifanov and B.~Shwartz, for providing their data.  This work was
supported in part by the Ministerio de Ciencia e Innovaci\'on under
grant CICYT-FEDER-FPA2008-01430, the EU Contract
No.~MRTN-CT-2006-035482, ``FLAVIAnet'', the Spanish Consolider-Ingenio
2010 Programme CPAN (CSD2007-00042), and the Generalitat de Catalunya
under grant SGR2009-00894.

%







\begin{thebibliography}{00}






 \bibitem{LR84} H.~Leutwyler and M.~Roos,
  Z.\ Phys.\  C {\bf 25} (1984) 91.


\bibitem{JOP4} M.~Jamin, J.~A.~Oller and A.~Pich,
  Phys.\ Rev.\  D {\bf 74}, (2006) 074009 
  [arXiv:hep-ph/0605095].


  \bibitem{Antonelli10}
  M.~Antonelli {\it et al.}, Eur. Phys. J. C, published online (2010),
  arXiv:1005.2323 [hep-ph].


\bibitem{Belle}  D.~Epifanov {\it et al.}  [Belle Collaboration],
  Phys.\ Lett.\  B {\bf 654} (2007) 65 
  [arXiv:0706.2231 [hep-ex]].


\bibitem{Babar} 
S.~Paramesvaran [BaBar Collaboration], proceedings of Meeting of the
Division of Particles and Fields of the American Physical Society (DPF
2009), Detroit, Michigan, 26-31 Jul 2009, arXiv:0910.2884 [hep-ex].


\bibitem{BESIII} D.~M.~Asner et al,``Physics at BES-III'', Edited by K.~T.~Chao and Y.~F.~Wang, Int. J.
of Mod. Phys. A {\bf 24} Supplement 1, (2009) [arXiv:0809.1869].

\bibitem{BEJ}  D.~R.~Boito, R.~Escribano and M.~Jamin,
  Eur.\ Phys.\ J.\  C {\bf 59}, (2009) 821
  [arXiv:0807.4883 [hep-ph]];
  PoS  {\bf EFT09}  (2009) 064
[arXiv:0904.0425 [hep-ph]]

\bibitem{BEJ2010} D.~R.~Boito, R.~Escribano and M.~Jamin,
  JHEP {\bf 1009} (2010) 031
  [arXiv:1007.1858 [hep-ph]].


\bibitem{Lattice}G.~Colangelo,  talk presented at EuroFlavour 2010, Munich, Germany. 


\bibitem{Omnes} R.~Omn\`es,
  Nuovo Cim.\  {\bf 8} (1958) 316.



\bibitem{PP2001}A.~Pich and J.~Portol\'es,
  Phys.\ Rev.\  D {\bf 63} (2001)  093005 
  [arXiv:hep-ph/0101194].



\bibitem{Bernardetal}
 V.~Bernard, M.~Oertel, E.~Passemar and J.~Stern,
  Phys.\ Lett.\  B {\bf 638}(2006)  480  
  [arXiv:hep-ph/0603202];
  Phys.\ Rev.\  D {\bf 80} (2009) 034034  
  [arXiv:0903.1654 [hep-ph]].





\bibitem{JPP20062008} M.~Jamin, A.~Pich and J.~Portol\'es,
  Phys.\ Lett.\  B {\bf 640} (2006) 176 
  [arXiv:hep-ph/0605096];
  Phys.\ Lett.\  B {\bf 664} (2008) 78 
  [arXiv:0803.1786 [hep-ph]].


\bibitem{GL}J.~Gasser, H.~Leutwyler,
  Nucl.\ Phys.\  {\bf B250} (1985) 517.


\bibitem{JOP1} M.~Jamin, J.~A.~Oller and A.~Pich,
  Nucl.\ Phys.\  B {\bf 622} (2002) 279 
  [arXiv:hep-ph/0110193].



\bibitem{LASS} D.~Aston {\it et al.},
  Nucl.\ Phys.\  B {\bf 296} (1988) 493.


\bibitem{Estabrooks} P.~Estabrooks, R.~K.~Carnegie, A.~D.~Martin, W.~M.~Dunwoodie, T.~A.~Lasinski and D.~W.~G.~Leith,
  Nucl.\ Phys.\  B {\bf 133} (1978) 490.

 \end{thebibliography}


\end{document}